\documentclass[pre,showpacs,twocolumn]{revtex4}
\usepackage{graphicx} % Include figure files
\usepackage{dcolumn}  % Align table columns on decimal point
\usepackage{bm}       % bold math
\usepackage{amsmath}
\usepackage{amsfonts}
\usepackage{amssymb}

\begin{document}

\title{Strong far field coherent scattering of ultraviolet radiation
  by holococcolithophores.}

\author{R. Quintero-Torres and J.L. Arag\'on} \affiliation{Centro de
  F\'{\i}sica Aplicada y Tecnolog\'{\i}a Avanzada, Universidad
  Nacional Aut\'onoma de M\'exico, Apartado Postal 1-1010, Quer\'etaro
  76000, M\'exico.}

\author{M. Torres} \affiliation{Instituto de F\'{\i}sica Aplicada,
  Consejo Superior de Investigaciones Cient\'{\i}ficas, Serrano 144,
  28006 Madrid, Spain.}

\author{M. Estrada and L. Cros} \affiliation{Institut de Ci\`encies
  del Mar, Centre Mediterrani d'Investigacions Marines i Ambientals,
  Consejo Superior de Investigaciones Científicas, Passeig Marítim de
  la Barceloneta 37-49, 08003 Barcelona, Spain.}

\begin{abstract}
  By considering the structure of holococcoliths (calcite plates that
  cover holococcolithophores, a haploid phase of the coccolithophore
  life cycle) as a photonic structure, we apply a discrete dipolar
  approximation to study the light backscattering properties of these
  algae. We show that some holococcolith structures have the ability
  to scatter the ultraviolet (UV) radiation. This property may
  represent an advantage for holococcolithophores possessing it, by
  allowing them to live higher in the water column than other
  coccolithophores.
\end{abstract}

\maketitle

\section{Introduction}

Recently, attention is paid to photonic structures in biology as
living physical systems generated by evolution (Ref. \cite{Vukusic}
and references therein). Coccolithophores are unicellular algae
distinguished by a covering of calcium carbonate plates, called
coccoliths, of still unknown purpose in spite of numerous hypotheses
about their possible functions
\cite{Braarud,Young,Thierstein}. Coccolithophores are found throughout
the euphotic zone of the sea and constitute a significant fraction of
the phytoplankton in open ocean environments. They impact greatly on
marine ecosystems, and play an important role in the global carbon
cycle \cite{Westbroek}.The optical effects of coccolithophores have
been widely investigated \cite{Holligan}, but their exact optical
properties have not yet been determined. It is known that coccoliths
do not absorb light but rather reflect or scatter it by acting as tiny
mirrors dispersed in the water \cite{Tyrrel}; consequently, the main
optical impact of coccolithophores is an increase in light
scattering. The light backscattering properties of oceanic mesoscale
coccolithophore blooms have been measured experimentally. It has also
been observed that coccolithophore blooms cause a large increase of
the ocean albedo at all light wavelengths but it is particularly
outstanding in the threshold of the UV spectrum \cite{Balch,Tyrrel}.
The Mie model has been used to fit the experimental measurements,
under the assumption that the coccolithophore cell covering
(coccosphere) could be represented as a homogeneous sphere
\cite{Vaillancourt}.

In the present work, we turn our attention to holococcoliths, special
coccoliths that have a periodic structure of calcite crystallites and
form the cell covering of holococcolithophores (the haploid phase of
many coccolithophores). We present a dipolar multiscattering model
that considers an ensemble of dielectric calcite nanospheres, arranged
according to the structure of some holococcoliths. A dipolar
multiscattering model is suitable after considering the size and shape
of the coccolith microstructure and the involved frequency
range. Neither the diffraction theory can be successfully applied at
this dimensions nor the classical optics on calcitic microlenses
\cite{Alzenberg}. In particular, as shown in Figure \ref{fig:fig1}, we
will consider holococcoliths that present a triangular layer on top of
an hexagonal one of tiny similar-sized calcite crystalline nanospheres
(the so-called crystallites). As we shall show, these crystalline
structures enhance the light scattering for UV radiation and thus
constitute natural calcitic photonic structures.

In the top layers of aquatic environments, phytoplankton receive solar
energy, necessary to drive photosynthesis, but are simultaneously
exposed to UV radiation that can affect biological processes and
damage DNA and other cell compounds \cite{Hader,Day,Buma,Garde}. A
widespread phytoplankton response is the production of sunscreening
compounds such as mycosporine-like amino acids \cite{Morrison}. With
regard to coccolithophores, it has been suggested that coccoliths
could exert a protective effect by reflecting UV light
\cite{Young}. According to our results, the crystalline photonic
structure of certain holococcoliths enhances UV backscattering. This
could have some ecological advantages and could represent an evolutive
adaptation of some holococcolithophores, implying a particular choice
of the crystalline structure parameters, \emph{i.e.} cell parameter
and calcite nanosphere radius, of the holococcoliths.

\section{The model}

\begin{figure}[t!]
\begin{center}
  \includegraphics[width=5.0cm]{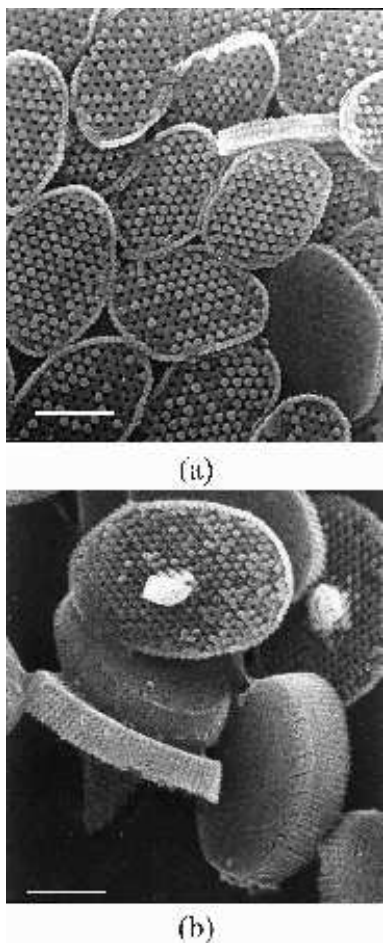}
  \caption{(a) \emph{Calcidiscus leptoporus} HOL (formerly
    \emph{Crystallolithus rigidus}). (b) \emph{Helicosphaera carteri}
    HOL (formerly \emph{Syracolithus catilliferus}).  In both cases,
    images are taken from Ref. \cite{Cros2} (Figs. 62D and 10D,
    respectively) and the scale bar is 1 $\mu$m.} \label{fig:fig1}
\end{center}
\end{figure}

\begin{figure}[t!]
\begin{center}
 \includegraphics[width=7.0cm]{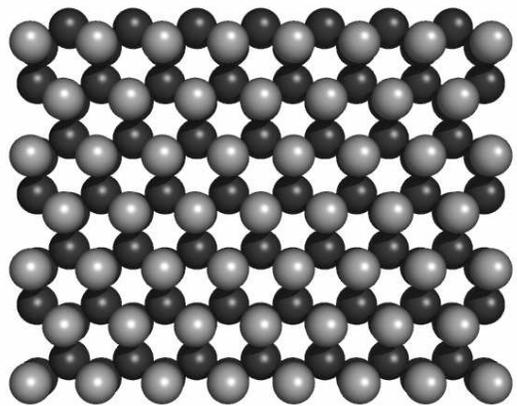}
 \caption{Model with spheres representing the two layers observed in
   holococcoliths as those shown in Figure \ref{fig:fig1}. Spheres
   with a radius of 0.045 $\mu$m are arranged in a plane of hexagons
   (dark spheres) with an edge-length of 0.09 $\mu$m, covered by a
   triangular layer (light gray spheres) with an edge-length of
   0.16 $\mu$m. The separation between planes is
   0.09 $\mu$m.} \label{fig:fig2}
\end{center}
\end{figure}

In Figure \ref{fig:fig1} images of holococcoliths from
\emph{Calcidiscus leptoporus} HOL (formerly \emph{Crystallolithus
  rigidus}) and \emph{Helicosphaera carteri} HOL, formerly
\emph{Syracolithus catilliferus} are shown. Inspired in the observed
hexagonal layer of calcite crystals with a triangular layer on top, we
consider the model, shown in Figure \ref{fig:fig2}, that consists of
calcium carbonate spheres with a radius of 0.045 $\mu$m, arranged in a
plane of hexagons with an edge-length of 0.09 $\mu$m, covered by a
triangular layer with an edge-length of 0.16 $\mu$m (triangular and
hexagonal layers are 0.09 $\mu$m apart). All this structure is immersed
in water with a refractive index of 1.333 at a wavelength of 633 nm. 
Our model assumes a holococcolith surface of 2 ($\mu$m)$^2$.

% \begin{figure}[t!]
%   \centering
%   \subfloat[][]{\includegraphics[width=5.0cm]{figure1a.eps}} \qquad
%   \subfloat[][]{\includegraphics[width=5.0cm]{figure1b.eps}}
%   \caption{(a) \emph{Calcidiscus leptoporus} HOL (formerly
%     \emph{Crystallolithus rigidus}). (b) \emph{Helicosphaera carteri}
%     HOL (formerly \emph{Syracolithus catilliferus}).  In both cases,
%     images are taken from Ref. \cite{Cros2} (Figs. 62D and 10D,
%     respectively) and the scale bar is 1 $\mu$m.} \label{fig:fig1}
% \end{figure}

Based on the size and shape of the holococcolith microstructure, we
adopt a discrete multipolar approximation to study radiation
scattering. In this approximation, the target is replaced by an array
of point dipoles or, in general, multipoles, which become
electromagnetic scatterers. In each vertex, the polarization of the
incident radiation field induces an electromagnetic multipole that
oscillates with a specific phase, defined by its position in space,
and radiates energy in all directions. Far away from the multipole,
the angular distribution of the radiation scattered by this structure
is given by the coherent superposition of the emission of each
individual source, measured in the test point in space. Thus, in our
model, the calcium carbonate spheres arranged in the hexagonal and
triangular layers are replaced by point dipoles.

The power radiated in the direction ${\bf n}$, with polarization
$\boldsymbol\epsilon$, per unit of solid angle, that results from the
incident radiation in the direction ${\bf n}_0$, with polarization
$\boldsymbol\epsilon _0$ is the derivative of the scattering cross
section $\sigma$ with respect to the solid angle $\Omega$
\cite{Jackson}:

\begin{equation}
  \frac{d \sigma}{d \Omega} = \frac{k ^4}{\left( 4 \pi \epsilon_0 E_0 \right) ^2} 
  \left| \sum_j \left[ \boldsymbol\epsilon ^* \cdot {\bf p}_j \right] 
    e ^{i {\bf q} \cdot {\bf r}} \right| ^2,
\end{equation}

Where ${\bf q} = k ({\bf n}_0 - {\bf n})$, $\lambda$ is the
wave-length, $k = 2 \pi n / \lambda$, $E_0$ is the external electric
field and ${\bf p}_j$ are the dipolar moments. The sum extends over
all nano-spheres. The calculation was carried out assuming that the
modeled arrangement has an arbitrary orientation with respect to the
incidence direction with the angles ($\theta$, $\phi$) for the
standard spherical coordinates. The polarization of the incident
electromagnetic field is well defined; both $S$ and $P$ polarization
are considered.

\section{Results}

\begin{figure}[t!]
\begin{center}
 \includegraphics[width=9.0cm]{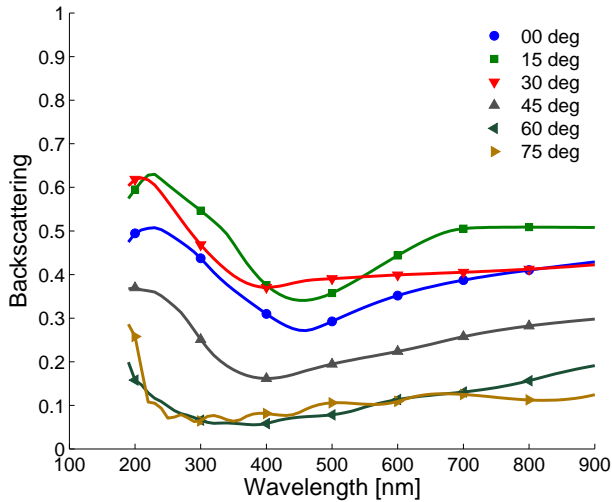}
 \caption{(Color online) Radiation backscattering of the periodic structure 
  shown in Fig. \ref{fig:fig2}.} \label{fig:fig3}
\end{center}
\end{figure}

In Figure \ref{fig:fig3} the backscattering intensity against
radiation wave-length is shown for different azimuthal angle incidence
($\theta$). Negligible differences between $P$ and $S$ polarization
are measured. We present here results for $S$-polarization of the
incident radiation.

As can be seen in Figure \ref{fig:fig3}, the geometrical structure in
Fig. \ref{fig:fig2} \textquotedblleft sees\textquotedblright ~ short wavelengths
 better than the large ones. For non-grazing incidences the backscattering 
 is strongly enhanced in the UV range (radiation wavelength less than 400 nm).

With the aim to ascertain the importance of the geometrical
arrangement, we compared the optical properties of a hexagonal and an
amorphous structure. First, we modeled a geometrical structure using
the hexagonal and triangular layers as in Fig. \ref{fig:fig2}. Next,
the position of each scatterer was scrambled in a random manner inside
a box of 2.54$\times$1.54$\times$0.2 $\mu$m. This yields a random
(amorphous) structure of dielectric nanospheres (dipoles)

As compared with an amorphous distribution of calcitic spheres, the
present case of periodic location for the scatters shows a sharp
variation for both wavelength and orientation.

\begin{figure}[t!]
\begin{center}
 \includegraphics[width=9.0cm]{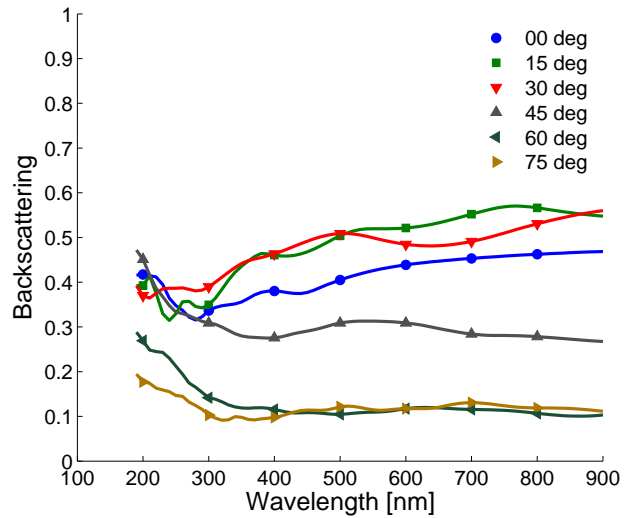}
 \caption{(Color online) Radiation backscattering of the amorphous structure.}
\label{fig:fig4}
\end{center}
\end{figure}

In an amorphous arrangement (Fig. \ref{fig:fig4}), the different plots
represent different orientations with respect to the incident
electromagnetic field. By adding all the backscattered radiation, we
can see an even effect at all wavelengths, with small variation for
different orientations. The periodic structure shows a sharp variation
for both wavelength and orientation. It also filters more efficiently
radiation wavelengths below 400 nm and the total amount of radiation
backscattered is clearly greater for this case that for the amorphous
one.

\section{Discussion}
The arrangement of the calcite crystallites in hexagonal or other
geometric patterns is usual in holococcolithophores
(ex. \emph{Syracosphaera pulcra} HOL (formerly \emph{Calyptrosphaera
  oblonga}), \emph{Calyptrolithophora papillifera},
\emph{C. gracillima}, \emph{Holococcolithophora heimdaliae},
\emph{H. dentata}). Among them, a few ones present two or more clearly
defined crystallite layers (for example, the here modeled
\emph{Calcidiscus leptoporus} HOL and \emph{Helicosphaera carteri}
HOL, formerly \emph{Syracolithus catilliferus}). Some other
holococcolithophores present a clear hexagonal layer and a careful
analysis shows an incipient triangular layer (with an identical
arrangement as those shown in Fig. 2) which, we hypothesize, is
necessary to scatter the UV light. This is the case, for instance, of
\emph{Calyptrolithophora papillifera} and \emph{Syracosphaera pulchra}
HOL, formerly \emph{Calyptrosphaera oblonga} HOL \cite{Cros2}.

The results produced with the proposed coccolith geometry help to
understand published work regarding the interaction between radiation
and this complex structure. The simulations for the modeled hexagonal
and triangular pattern (Figure \ref{fig:fig3}) reveal that the UV
light can be more strongly backscattered than that of other
wavelengths. At the same time, the decreased backscattering around 
400-700 nm would minimize the loss of photosynthetically active light. 
It has been shown \cite{Cros} that holococcolithophores tend to be found 
higher in the water column than heterococcolithophores, the diploid 
phase of the coccolithophore life cycle, which have a different coccolith 
organization. The two-layered coccolith structure of holococcolithophores 
such as those studied here suggests a possible strategy of increasing 
the reflection of UV light away from the cell, thus enhancing the 
ability of the organism to live higher in the water column, as it 
has been speculated \cite{Young} with respect a potential light regulation 
function of the coccoliths. The differential backscattering of UV light 
may represent an additional adaptation for the utilization of different 
ecological niches by cells with diverse coccolith structures.

\begin{acknowledgments}
  Computational support from B. Mill\'an is gratefully
  acknowledged. This work has been partially supported by DGAPA-UNAM
  (Grant Nos. IN-117806 and IN-118406), CONACyT (Grant No. D40615-F), 
  the Spanish MCYT (Grant No. FIS2004-03237), EU project FMRX-CT97-0113 
  and MARPLAN, a Responsive Mode Project within the MarBEF EU Network 
  of Excellence, funded under the Sixth Framework Programme of the EU.
\end{acknowledgments}


\begin{thebibliography}{10}
\bibitem{Vukusic} P. Vukusik and J.R. Sambles, Nature (London)
  \textbf{424}, 852 (2003).

\bibitem{Braarud} T. Braarud, K. R Gaarder, J. Markali and E. Nordli,
  Nytt Mag. Bott. \textbf{1}, 129 (1952).

\bibitem{Young} J.R. Young, in \emph{Coccolithophores}, edited by
  A. Winter and W.G. Siesser (Cambridge University Press, Cambridge,
  1994) p.63.

\bibitem{Thierstein} H.R. Thierstein and J.R. Young (Editors),
  \emph{Coccolithophores. From Molecular Processes to Global Impact}
  (Springer Verlag, New York, 2004).

\bibitem{Westbroek} P. Westbroek, J. E. van Hinte, G.-J. Brummer,
  M. Veldhuis, C.  Brownlee, J. C. Green, R. Harris and B. R. Heimdal,
  in \emph{The Haptophyte Algae}, edited by J. C. Green and
  B. S. C. Leadbeater, Systematics Association Special Volume 51
  (Clarendon Press, Oxford, 1994), p. 321.

\bibitem{Holligan} P.M. Holligan and W.M. Balch, in \emph{Particle
    Analysis in Oceanography}, edited by S. Demers, Proc. NATO ASI
  Ser., Ser. G \textbf{27}, 301 (1991).

\bibitem{Tyrrel} T. Tyrrell, P.M. Holligan and C.D. Mobley,
  J. Geophys. Res. \textbf{104}, 3223 (1999).

\bibitem{Balch} W.M. Balch, K.A. Kilpatrick, P. Holligan. D. Harbour
  and E. Fernandez, Limnol. Oceanogr. \textbf{41}, 1684 (1996).

\bibitem{Vaillancourt} R.D. Vaillancourt, C.W. Brown, R.R.L. Guillard
  and W.M. Balch, J. Plankton Res. \textbf{26}, 191 (2004).

\bibitem{Alzenberg} J. Alzanberg, A. Tkachenko, S. Welner, L. Addali
  and G. Hendler, Nature (London) \textbf{412}, 819 (2001).

\bibitem{Hader} D.-P. H\"ader, H.D. Kumar, R.C. Smith and R.C. Worrest, 
  J. Photoch. Photobio. B \textbf{46}, 53 (1998).

\bibitem{Day} T.A. Day and P.J. Neale, Annu. Rev. Ecol. Syst. \textbf{33}, 
  371 (2002).

\bibitem{Buma} A.G.J. Buma, T. van Oijen, W. van de Poll, M.J.W. Veldhuis, 
  and W.W.C. Gieskes, J. Phycol. \textbf{36}, 296 (2000).

\bibitem{Garde} K. Garde, and C. Cailliau, J. Exp. Mar. Biol. Ecol. 
  \textbf{247}, 99 (2000).

\bibitem{Morrison} J.R. Morrison and N.B. Nelson, Limnol. Oceanogr. 
  \textbf{49}, 215 (2004).

\bibitem{Cros2} L. Cros and J.L. Fortu\~no, Sci. Mar. \textbf{66}, 1
  (2002).

\bibitem{Jackson} J.D. Jackson, \emph{Classical Electrodynamics}
  (Wiley, U.S.A, 1999).

\bibitem{Cros} L. Cros, \emph{Planktonic coccolithophores of the NW
    Mediterranean}. Published Ph. D. Thesis (Publicacions de la
  Universitat de Barcelona, Barcelona, 2002).
\end{thebibliography}
\end{document}